# Theory of Scanning Tunneling Microscopy Probe of Impurity States in a $D$-Wave Superconductor

M.I. Salkola[1,*], A.V. Balatsky[2], and D.J. Scalapino[3]

[1] *Department of Physics and Astronomy, McMaster University, Hamilton, Ontario L8S 4M1, Canada*
[2] *Theoretical Division, Los Alamos National Laboratory, Los Alamos, New Mexico 87545*
[3] *Department of Physics, University of California, Santa Barbara, California 93106*
(February 22, 1996)

Scanning tunneling microscopy can provide a probe for the detailed study of quasiparticle states in high-$T_c$ superconductors. We propose that it can also be used to acquire specific information about impurity-induced quasiparticle states and the superconducting order-parameter structure. In particular, the local density of states is found to be sensitive to impurity-induced resonances and to the symmetry of the order parameter.

There has been much recent interest in the role of imperfections in unconventional superconductors, because they may provide a fruitful direction for understanding the pairing mechanism in high-$T_c$ superconductors. Nonmagnetic impurities are known in conventional superconductors to perturb the superconducting properties only weakly [1]. However, in superconductors with a nontrivial order-parameter phase, the same impurities are shown to be pair-breakers [2] producing a finite lifetime for quasiparticles near the nodes and a finite density of states at low energies. Impurities, therefore, provide practical and stringent tests for theoretical paradigms describing high-$T_c$ superconducting materials, and the lack of consensus regarding the proper identification of the pairing state provides a clear impetus for studying impurity effects further in more general situations.

Questions concerning spatially inhomogeneous quasiparticle states generated by impurities were first explored in the 60's [3]. It was shown that a magnetic impurity in an $s$-wave superconductor, interacting with the spin density of conduction electrons, produces a bound state inside the energy gap. In general, the overlap with the particle-hole continuum only allows virtual states to be formed with finite lifetimes. Similarly, it has been recently emphasized that, in a $d$-wave superconductor, a single scalar impurity generates a low-energy resonance which, in the unitary limit, becomes a marginally bound state of a highly anisotropic character [4].

In this Note, we study the effect of strongly-scattering impurities on quasiparticle states at low energies in a $d$-wave superconductor and discuss how these modified states could be observed in scanning tunneling microscopy (STM) which gives access to a much richer set of diagnostic tools by probing the local density of states [5]. To this end, we present a detailed description of the density of states as a function of space and energy near an impurity. Our main results are twofold: first, we emphasize that a scalar impurity breaks locally the particle-hole symmetry leading to experimentally observable consequences and, second, we show that the resonance behavior arising from the virtual impurity-bound state strongly enhances the local density of states at the resonance energy in the vicinity of an impurity. This enhancement is sensitive to the order-parameter structure, and its anisotropic spatial dependence reflects the orbital symmetry of the pairing state. Because the local density of states can be determined experimentally by measuring the local $I(V)$ characteristics, we suggest that STM studies can also reveal the nontrivial impurity-induced states in $d$-wave and other unconventional superconductors.

The local density of states of a $d$-wave superconductor near an impurity has been studied earlier both in the quasiclassical theory [6] and in the Born approximation where the impurity scattering acts as a weak perturbation [7]. It was found that impurity scattering modifies the local density of states in a manner which may serve as a finger print of $d$-wave pairing. While weakly-scattering impurities do not produce a strong dependence on the phase of the pairing state, the resonance behavior of the virtual impurity states in the energy domain allows, for example, the phase-sensitive contribution of the pairing state to be isolated and explored. In addition to the Josephson effect, this provides a potentially attractive, new method for probing experimentally both the amplitude *and* the phase of the order parameter of high-$T_c$ superconductors [8].

Specifically, we consider a repulsive $\delta$-potential impurity in a superconductor with a real gap function $\Delta(\mathbf{k})$: $U(\mathbf{r}) = U\delta(\mathbf{r})$, $U > 0$; our results can easily be generalized for an attractive impurity. Scattering of quasiparticles from the impurity is described by a $T$-matrix, $\hat{T}(\omega)$, which is independent of wave vector for $s$-wave scattering. The Green's function in the presence of the impurity is $\hat{G}(\mathbf{r}, \mathbf{r}'; \omega) = \hat{G}^{(0)}(\mathbf{r} - \mathbf{r}', \omega) + \hat{G}^{(0)}(\mathbf{r}, \omega)\hat{T}(\omega)\hat{G}^{(0)}(-\mathbf{r}', \omega)$, where both $\hat{G}^{(0)}(\mathbf{r}, \omega)$ and $\hat{T}(\omega)$ are matrices in Nambu particle-hole-spinor space, with $\omega + i0^+$ implied hereafter. The Green's function of a clean system is $\hat{G}^{(0)}(\mathbf{k}, \omega) = [\omega\hat{\tau}_0 - \epsilon(\mathbf{k})\hat{\tau}_3 - \Delta(\mathbf{k})\hat{\tau}_1]^{-1}$, where $\epsilon(\mathbf{k})$ is the quasiparticle energy, $\hat{\tau}_\alpha$ ($\alpha = 1, 2, 3$) are the Pauli matrices, and $\hat{\tau}_0$ is the unit matrix in Nambu space.

For a system with $d$-wave pairing and a particle-hole symmetric normal-state energy spectrum, the $T$-matrix has only two nonzero components [9]: $\hat{T} = T_0\hat{\tau}_0 + T_3\hat{\tau}_3$. They are $T_0(\omega) = G_0(\omega)/[c^2 - G_0(\omega)^2]$ and $T_3(\omega) = c/[c^2 - G_0(\omega)^2]$, where the dimensionless parameter $c$ is inversely proportional to the strength of the impurity potential $U$: $c = (\pi N_F U)^{-1}$. The momentum-averaged Green's functions, $G_\alpha(\omega) \equiv G_\alpha^{(0)}(\mathbf{r} = 0, \omega)$, are conveniently expressed in terms of their real-space analogues $G_\alpha^{(0)}(\mathbf{r}, \omega) = \frac{1}{2\pi N_F} \int dk^2 \, \mathrm{Tr}\, \hat{\tau}_\alpha \hat{G}^{(0)}(\mathbf{k}, \omega) e^{i\mathbf{k}\cdot\mathbf{r}}$ ($\alpha = 0, \ldots, 3$); $N_F$ is the density of states per unit volume at the Fermi energy in the normal state. The virtual or bound states in the single-impurity problem are given by the poles of the $T$-matrix: $G_0(\Omega) = \pm c$. Choosing the gap function of $d_{x^2-y^2}$ symmetry at the Fermi surface so that $\Delta(\varphi) = \Delta_0 \cos 2\varphi$, we obtain

$$G_0(\omega) = -(2\epsilon/\pi)\left[\mathrm{sgn}(\omega) K(\sqrt{1-\epsilon^2}) + iK(\epsilon)\right], \quad (1)$$

for $\epsilon \equiv |\omega|/\Delta_0 < 1$. Here, $K$ is the complete elliptic integral of the first kind. In principle, the solution for the poles is complex, indicating the resonant nature of the virtual state. For small $c$, this is easily seen from

$$\Omega_0 \simeq c\Delta_0 \frac{\pi/2}{\ln(8/\pi c)}, \quad \Gamma \simeq \Omega_0 \frac{\pi/2}{\ln(8/\pi c)}, \quad (2)$$

which solve $\Omega \equiv \mp\Omega_0 - i\Gamma$ to logarithmic accuracy. However, as $c \to 0$, the resonance becomes arbitrarily sharp. For Born scattering, $c \gg 1$, no resonance structure is generated.

STM measurements can probe the local density of states by measuring the differential conductance $dI/dV$ as a function of bias voltage $V$. We focus on the density of states at zero temperature defined as $N(\mathbf{r}, \omega) = -\mathrm{Im}\, G_{11}(\mathbf{r}, \mathbf{r}; \omega)$. (Hereafter the density of states of a superconductor is normalized relative to $N_F$.) It is convenient to divide it into two parts: $N(\mathbf{r}, \omega) = N_0(\omega) + \delta N(\mathbf{r}, \omega)$, where $N_0(\omega) = -\mathrm{Im}\, G_0(\omega)$ is the uniform density of states of a clean superconductor and $\delta N(\mathbf{r}, \omega)$ is the impurity-induced contribution. For $|\omega| \ll \Delta_0$, $N_0(\omega) = |\omega|/\Delta_0$. Using the above results, the impurity-induced contribution can be written as:

$$\delta N(\mathbf{r}, \omega) = -\mathrm{Im}[G_+^{(0)}(\mathbf{r}, \omega) T_+(\omega) G_+^{(0)}(-\mathbf{r}, \omega) \\ + G_1^{(0)}(\mathbf{r}, \omega) T_-(\omega) G_1^{(0)}(-\mathbf{r}, \omega)], \quad (3)$$

where we have defined $G_+^{(0)} = G_0^{(0)} + G_3^{(0)}$ and $T_\pm = T_0 \pm T_3$. In the limit $c \to \infty$ (i.e., $U \to 0$), this equation reduces to the Born form [7], where $T_0 = \mathcal{O}(c^{-2})$ and $T_3 = \mathcal{O}(c^{-1})$. However, in the opposite limit, the resonance behavior becomes an important new ingredient in the problem, as can be seen by noting that

$$T_\pm(\omega) = -\frac{1}{G_0(\omega) \mp c}. \quad (4)$$

Thus, in the neighborhood of $\omega = \pm\Omega_0$, a rapid modulation of the differential conductance is expected. Depending on the relative magnitudes of the resonance energy $\Omega_0$ (determined by $c$) and the probe energy $\omega$ (i.e., the bias voltage), three regimes of interest appear. Here, for brevity, we consider only the strong-scattering limit, for which $\Omega_0 \ll \Delta_0$. (i) For small energies $|\omega| \ll \Omega_0$, the problem reduces to that of a weakly scattering impurity, described by the Born approximation [7]. In this case, $T_+ \simeq -T_-$. (ii) In the opposite limit, where $|\omega| \gg \Omega_0$, we have $T_+ \simeq T_-$. (iii) In the cross-over regime, $|\omega| \sim \Omega_0$, they are rapidly varying functions of $\omega$. In fact, $T_\pm$ has a resonance at $\omega = \mp\Omega_0$. This resonance is important already at a moderately small value of $c$. As $c$ decreases, the amplitude of the resonance peak in $\mathrm{Im}\, T_\pm$ increases rapidly and, at the same time, the resonance becomes sharper but this happens only at a much slower rate as is evident from Eq. (2).

Recall the three characteristic scales of length: the impurity-induced length $\ell_0 = \hbar v_F / \Omega_0$, the coherence length $\xi_0 = \hbar v_F / \Delta_0$, and the Fermi length $\ell_F = 2\pi/k_F$, where $v_F$ is the Fermi velocity and $k_F$ is the Fermi wave vector. The applicability of the continuum theory and for having a pronounced impurity-induced resonance require that $\ell_F \ll \xi_0 \ll \ell_0$. The nature of the local density of states depends on the ratio of the distance $r$ from the impurity to these scales. Below, we focus on $N(\mathbf{r}, \omega)$ in the vicinity of the impurity because, for $r > \ell_0$, the intensity of the STM signal is rapidly saturating function of $r$, and the tunneling current will be essentially spatially uniform.

(i) *A d-wave superconductor*. First, consider the differential conductance at the repulsive impurity site, $\mathbf{r} = 0$ (i.e., $r \ll \ell_F$). Because $\hat{G}^{(0)}(\mathbf{r} = 0, \omega) = G_0(\omega)\hat{\tau}_0$, the density of states has a simple form: $N(\mathbf{r} = 0, \omega) = -\mathrm{Im}[cG_0(\omega)/(c - G_0(\omega))]$. For finite $c$, the van Hove singularity in the original density of states is replaced by zero at $\omega = \pm\Delta_0$. Overall, for a weak impurity, $c \gg 1$, the density of states is approximately the same as for a clean superconductor: $N(\mathbf{r} = 0, \omega) = N_0(\omega) + \mathcal{O}(c^{-1})$. The leading-order impurity effect is the spectral-weight transfer within the energy gap region ($|\omega| < \Delta_0$) from positive to negative energies. In contrast, for a strong impurity, $c \ll 1$, the situation becomes much more interesting. For $|\omega| \gg \Omega_0$, the density of states is negligible: $N(\mathbf{r} = 0, \omega) = \mathcal{O}(c^2)$, whereas, for $|\omega| \ll \Omega_0$, we recover the unperturbed density of states. In the crossover regime and for $\omega < 0$, $N(\mathbf{r} = 0, \omega)$ has a clear peak of magnitude approximately proportional to $c \log c^{-1}$, whereas, for $\omega > 0$, it shows only a very shallow feature; see Fig. 1. As $c \to 0$, the weight of both of these features vanishes continuously, and their position scales with $\Omega_0$ to zero energy. The resonance peak at $\omega < 0$ persists up to $c \sim 1$ where it appears as a broad structure. It is natural to interpret this peak as a reflection of the impurity-induced quasiparticle states, which are hole(electron)-like for a repulsive (attractive) potential. Consequently, an STM experiment will see only a small differential conductance at low energies ($|\omega| \sim \Omega_0$), and approximately zero conductance at higher energies. The depletion of the quasiparticle states is a result of a strong impurity potential which allows quasiparticles to be excited essentially only at very high energies $|\omega| \gg \Delta_0$ [10]. Because particles predominantly of hole ($c > 0$) or electron ($c < 0$) character can reside at the impurity site, a strongly-scattering impurity leads simultaneously to a nearly complete local pair-breaking effect. In the unitary limit ($c \to 0$) and at finite energies ($|\omega| \ll |U|$), the density of states vanishes for both signs of the impurity potential in both the electron and hole channels.

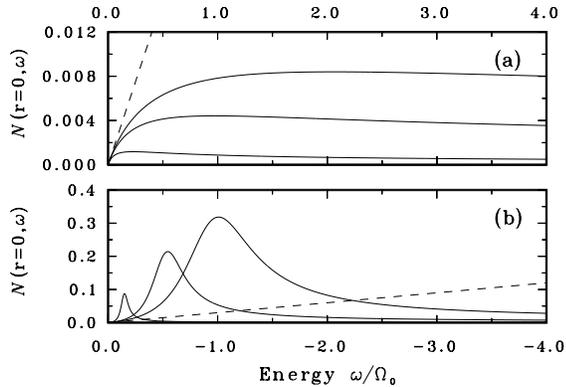

FIG. 1. The local density of states $N(\mathbf{r}=0,\omega)$ of a $d$-wave superconductor for (a) $\omega > 0$ and (b) $\omega < 0$ at the impurity site as a function of energy $\omega$ for $c = 0.02$, $0.06$, and $0.10$ (from bottom to top). Here, $\Omega_0 = 0.03\Delta_0$ is the impurity resonance energy for $c = 0.10$. The density of states $N_0(\omega) = |\omega|/\Delta_0$ of a clean $d$-wave superconductor is shown by the dashed line. Note the different density scales.

Next, consider the density of states in the neighborhood of the impurity, $\ell_F \lesssim r < \ell_0$, the focus being at distances $r \sim \xi_0$ and energies $|\omega| \sim \Omega_0 \ll \Delta_0$. Thus, in computing the Green's functions, we may neglect contributions proportional to $\omega/\Delta_0$, because they become important only at large distances $r > \ell_0$ or at higher energies [11]. Along the gap extrema, their dominant part can be estimated as $G_{1,+}^{(0)}(\mathbf{r},\omega) \sim (\ell_F/r)^{1/2} e^{-r/\xi_0} + \mathcal{O}(\ell_F/r)$. Clearly, along the diagonal directions where the energy gap has a line of nodes, $G_1^{(0)}$ vanishes. In contrast, $G_+^{(0)}(\mathbf{r},\omega) \sim \xi_0/r$. This result, valid for distances $r \lesssim \ell_\omega \equiv \hbar v_F/|\omega|$, is the leading contribution in $\hat{G}^{(0)}(\mathbf{r},\omega)$, for $|\omega| \sim \Omega_0$ and $\xi_0 < r < \ell_0$. This type of behavior, but extending to arbitrary long distances and reflecting the impurity-induced marginally bound state [4], can be observed only in the unitary limit where we first take $c \to 0$ and then $\omega \to 0$, because $\text{Im}\,T_\pm(0) = 0$, for $c \neq 0$.

A number of important and novel results follow from the qualitative behavior of $T_\pm$ and $G_{1,+}^{(0)}$. First, it is in-triguing to note that the impurity-induced STM signal is most strongly enhanced in the horizontal and vertical directions (the extrema directions of the $d$-wave energy gap) at distances $r \ll \xi_0$ from the impurity when the bias voltage passes through the resonance condition $\omega = \pm\Omega_0$. The STM signal, proportional to $N(\mathbf{r},\omega)$ and shown in Fig. 2, has an unusually large maximum at $r \sim \ell_F$ along these directions. Second, further away from the impurity ($r \sim \xi_0$), the situation is reversed: the STM signal is enhanced along the diagonals for $\omega = -\Omega_0$ (the hole-like resonance) and in the directions close to the diagonals for $\omega = \Omega_0$ (the electron-like resonance). For $r \lesssim \xi_0$, this enhancement should be large enough to be clearly observable, as illustrated in Fig. 3. However, at large distances, the impurity-induced contribution $\delta N(\mathbf{r},\pm\Omega_0)$ decays as $r^{-2}$ making its detection experimentally demanding. Third, particle-hole symmetry of the local density of states is manifestly broken by the impurity. Finally, our results are naturally understood by the observation that, for $c \ll 1$, the problem effectively reduces into an isolated impurity and the superconductor with a vacancy.

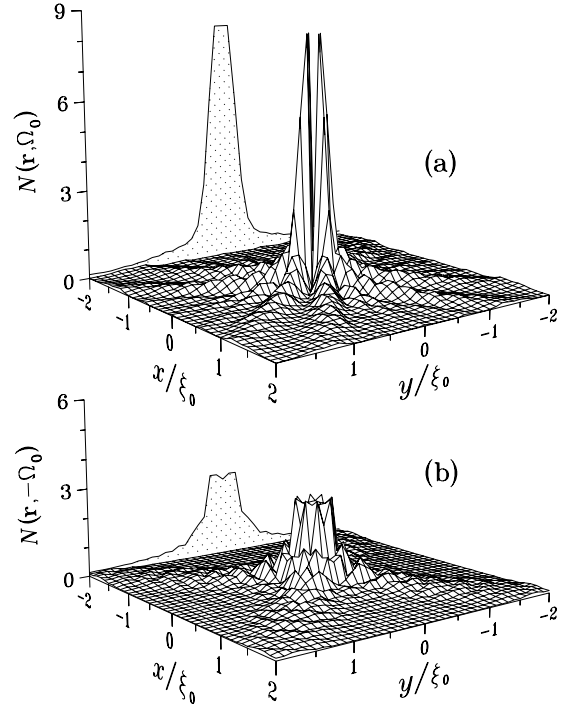

FIG. 2. The local density of states $N(\mathbf{r},\omega)$ of a $d$-wave superconductor for (a) $\omega = \Omega_0$ and (b) $\omega = -\Omega_0$ as a function of position $\mathbf{r} = (x,y)$ around a repulsive impurity ($c = 0.1$) placed at $\mathbf{r} = 0$. The impurity resonance energy is $\Omega_0 = 0.03\Delta_0$ and $\xi_0/\ell_F = 6$. The maximum values of $N(\mathbf{r},\pm\Omega_0)$ are approximately 8.3 and 2.9 (in units of $N_F$). The shaded profile shows the projection of $N(\mathbf{r},\pm\Omega_0)$ on the vertical plane.

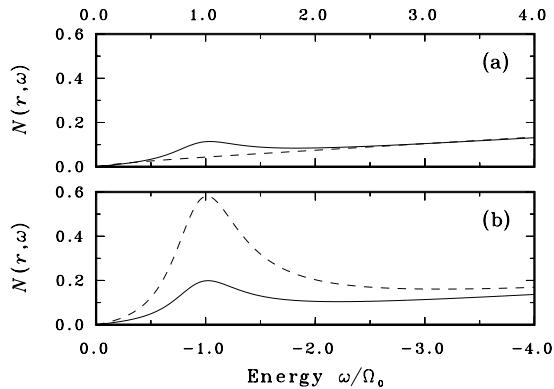

FIG. 3. The local density of states $N(\mathbf{r}, \omega)$ of a $d$-wave superconductor for (a) $\omega > 0$ and (b) $\omega < 0$ as a function of energy $\omega$ at the distance $r/\xi_0 = 0.9$ along the horizontal/vertical (solid lines) and diagonal (dashed lines) directions from the repulsive impurity ($c = 0.1$). The impurity resonance energy is $\Omega_0 = 0.03 \Delta_0$ and $\xi_0/\ell_F = 6$.

This description leads to two impurity-induced features that simultaneously exist in the system: a state associated with the impurity site itself and located at a very high energy ($\sim U$), and a scattering resonance induced by the vacancy and describing a low-energy ($\Omega_0 \ll \Delta_0$) "surface" state whose maximum weight is in the adjacent sites to the defect. For non-zero $c$, these two features are weakly coupled rationalizing the resonance in $N(\mathbf{r} = 0, \omega)$ at $\omega = -\Omega_0$. The resonant response of $N(\mathbf{r} \neq 0, \omega)$ at $\omega = \pm \Omega_0$ is the signature of the "surface" state, the virtual impurity-bound quasiparticle state.

(ii) *An anisotropic s-wave superconductor.* As a prototypical example, consider the gap function $\Delta(\mathbf{k}) = \Delta_0 |\cos 2\varphi|$. Although it has the same bulk density of states as the $d$-wave superconductor, there are no resonance states regardless of the value of $c$ [12]. This is a simplified version of the Anderson's theorem [1], which essentially states that disorder averages the energy gap over the Fermi surface; here the impurity scattering leads to the same effect, resulting locally in an effective isotropic energy gap of order $\Delta_0$. The anisotropic $s$-wave superconductor behaves then like an ordinary isotropic $s$-wave superconductor regarding the resonance phenomenon.

(iii) *A mixed s- and d-wave superconductor.* Finally, consider a pairing state described by a gap function $\Delta(\mathbf{k}) = \Delta_0 \cos 2\varphi + \Delta_s$, where a small $s$-wave component is added to the $d$-wave state, $\Delta_s \ll \Delta_0$. This type of admixture of two angular-momentum states is relevant in a $d$-wave superconductor where the tetragonal symmetry is broken by an orthorhombic distortion. For $|\omega| \ll \Delta_0$, $G_1(\omega) \simeq \Delta_s/\Delta_0$, and in the $T$-matrix equations $c$ is replaced by an effective impurity strength $\tilde{c} = \sqrt{c^2 + (\Delta_s/\Delta_0)^2}$. This implies that the strength of the impurity potential is effectively cut off at large values limiting its ability to generate a quasiparticle resonance. Note that $G_1^{(0)}(\mathbf{r}, \omega)$ still approximately vanishes along the nodes of the energy gap, making the distorted $d$-wave case similar to that of the tetragonal $d$-superconductor so long as $\Delta_s/\Delta_0 < c$. Thus, a small $s$-wave component would be invisible in an STM experiment. Controlling the degree of the orthorhombicity, the position and the sharpness of the resonance can be tunned.

In summary, a central feature of low-energy quasiparticle states perturbed by an impurity is that they embody information about the underlying superconducting condensate and its symmetry. An important consequence of a strongly-scattering impurity is the quasiparticle resonance which is sensitive to the phase of the pairing state and could be detected by STM. This property offers an intriguing possibility for the identification of the structure of the pairing state in high-$T_c$ superconductors.

We are grateful to J.R. Schrieffer for helpful discussions on the subject. This work was supported in part by Natural Sciences and Engineering Research Council of Canada, the Ontario Center for Materials Research (M.I.S.), and by the U.S. Department of Energy (A.V.B. and D.J.S.). D.J.S. would like to acknowledge the hospitality of the program on Correlated Electrons at the Center of Material Science at LANL.